# Branching Ratios for the Three Most Intense Gamma Rays in the Decay of $^{47}$Ca

E. Paige Abel[1,2], Chloe Kleinfeldt[2], Colton Kalman[2], Gregory W. Severin[1,2]

[1]Department of Chemistry, Michigan State University, East Lansing, MI 48824, USA

[2]National Superconducting Cyclotron Laboratory, Michigan State University, East Lansing, MI 48824, USA

## Abstract

A sample of $^{47}$Ca produced through isotope harvesting at the National Superconducting Cyclotron Laboratory was used to measure branching ratios of 7.17(5)%, 7.11(5)%, and 75.0(5)% for the 489.2, 807.9, and 1297.1 keV characteristic gamma-rays, respectively. Based on these updated branching ratios, the ground state to ground state $^{47}$Ca to $^{47}$Sc beta decay branching ratio has been indirectly measured as 17.7(5)% and the ground state to 1297.1 keV excited state as 82.2(5)%. These values represent a greatly increased precision for all five branching ratios compared to the currently accepted values [1]. The measurements presented here were made relative to the ingrown $^{47}$Sc daughter in a $^{47}$Ca sample and the well-established 159.4 keV gamma-ray branching ratio and the half-life for the decay of $^{47}$Sc [2–4]. These measurements were supported by verifying that the half-lives measured from characteristic gamma-ray peaks over multiple spectra for both $^{47}$Ca and $^{47}$Sc were consistent with previously reported values. Additionally, the half-lives of both $^{47}$Ca and $^{47}$Sc were independently measured with Liquid Scintillation Counting to reverify the previously reported values used in this study to find updated gamma-ray branching ratio values.

## Introduction

$^{47}$Sc is a radionuclide of interest for theranostic nuclear medicine applications with a half-life on the order of days, a 100% low energy beta emission, a low energy 159.4 keV gamma-ray emission for SPECT imaging, and $^{43,44}$Sc as positron emitting diagnostic partners [5–8]. One way to produce high specific activity samples of $^{47}$Sc is through the production and subsequent decay of its parent radionuclide, $^{47}$Ca [5]. When producing and processing $^{47}$Ca, the branching ratios of its most intense characteristic gamma rays can be used to quantify $^{47}$Ca through High Purity Germanium (HPGe) spectroscopy. The accepted, evaluated values for these gamma-ray branching ratios as well as the two main $^{47}$Ca to $^{47}$Sc beta decay branching ratios have changed and increased in uncertainty from 1970 to 2007 [1,9]. For the quantification of $^{47}$Ca, the 19% uncertainty in the reported branching ratio of the 1297.1 keV gamma ray and the 20% uncertainty in that for both the 489.2 and 807.9 keV gamma rays lead to similarly large uncertainties in the absolute activity of $^{47}$Ca [1]. The change and uncertainty in these evaluated values has prompted the current remeasurement of the three main gamma rays in the decay of $^{47}$Ca to $^{47}$Sc.

One way to measure the branching ratios of the characteristic gamma rays in the decay of $^{47}$Ca is by utilizing the parent-daughter relationship between $^{47}$Ca and $^{47}$Sc. In a sample of these radionuclides where equilibrium has not yet been reached, the ingrowth of $^{47}$Sc relative to the decay of $^{47}$Ca can be used to quantify $^{47}$Ca and thereby measure the branching ratio of characteristic gamma rays of $^{47}$Ca. With these updated gamma-ray branching ratio values, a more precise value can be found for the ground state to ground state beta decay and the ground state to 1297 keV excited state beta decay of $^{47}$Ca to $^{47}$Sc.

In this work, the branching ratios for the three most intense characteristic gamma rays in the decay of $^{47}$Ca have been remeasured and used to find updated values for the two main $^{47}$Ca beta decay branching ratios. All branching ratio values reported here have increased precision over those currently accepted. Additionally, the half-lives of $^{47}$Ca and $^{47}$Sc have been remeasured using both HPGe and Liquid Scintillation Counting (LSC) and have been found to agree with those previously reported.

## Methods

*Production of Ca-47:*

The $^{47}$Ca used in this measurement was produced with a 140 MeV/nucleon $^{48}$Ca$^{20+}$ beam at the National Superconducting Cyclotron Laboratory (NSCL). This ion beam was implanted at an average intensity of 72 pnA for 9.8 hours in a flowing-water target, producing $^{47}$Ca amongst many other radionuclides. The flowing-water target was composed of a water-filled interior with a titanium alloy shell (Ti64: 6% aluminum, 4% vanadium, mass balanced with Ti). The product radionuclides formed in the target were transported through the flow of water from the target to a water system which was described in detail previously [10]. This system contained components to monitor and condition the water and collect product radionuclides from the water. Cationic radionuclides, such as $^{47}$Ca, were collected on a cation exchange resin bed (1.5 g resin, AG50W-X8, mesh size 20-50, BioRad) during and after irradiation to allow for maximum collection of $^{47}$Ca from the water. The process of implanting the $^{48}$Ca beam in the target, flowing water through the system, and collecting cations from the water is depicted schematically in Figure 1.

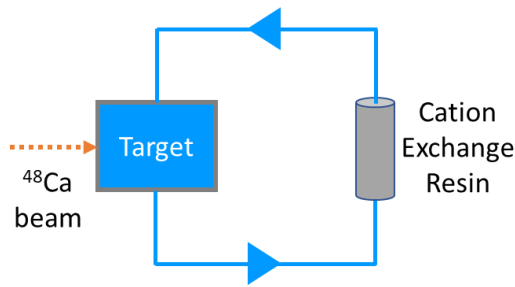

**Figure 1: Simplified Schematic of Target and Collection System**
The $^{48}$Ca beam is implanted in the flowing-water target, producing $^{47}$Ca that is transported to a cation exchange resin bed where it is adsorbed. The $^{47}$Ca is removed from the system on this resin bed and processed for further use.

*Purification of Ca-47:*

The purification of $^{47}$Ca collected from the water system is described in a recent publication [11] (See: *Separation Method 2: AG MP-50 with HCl*). In brief, the $^{47}$Ca was removed from the cation exchange resin bed in 2 M HCl. This solution was diluted to 0.2 M HCl to serve as the load solution for purification on a 2 g column of AG MP-50 (AG MP-50, 100-200 mesh size, BioRad). Following the load solution, rinse steps of 0.2 M HCl and 2 M HCl were used to remove other long-lived radionuclides that were produced in the water system and collected on the cation exchange resin bed in addition to $^{47}$Ca (*i.e.*, $^{7}$Be, $^{24}$Na,

$^{28}$Mg, $^{42,43}$K). Finally, 4 M HCl was used to elute purified $^{47}$Ca. Scandium radioisotopes ($^{44}$Sc, $^{44m}$Sc, $^{46}$Sc, $^{47}$Sc, $^{48}$Sc) which had accumulated on the cation exchange resin largely remained on the resin during the wash and elution steps. However, it is important to note that the scandium isotopes exhibited behavior which was consistent with that of hard hydrolyzable metal ions: they accumulated on many surfaces within the harvesting system and were difficult to control chemically. Therefore, the presence of Sc impurities in the purified $^{47}$Ca fraction could only be excluded at the detection limits of HPGe spectroscopy.

To produce a sample of $^{47}$Ca out of equilibrium with its daughter $^{47}$Sc and a purified $^{47}$Sc sample, the first step in a previously published pseudo generator for this parent-daughter pair was used [5]. About 10 mL of the 4 M HCl solution containing $^{47}$Ca was passed over 67 mg of DGA extraction chromatography resin (N,N,N',N'-tetra-n-octyldiglycolamide, normal resin, particle size 50-100 µm, TrisKem International) in a 1 mL ISOLUTE filtration column. The $^{47}$Sc that was generated in this solution was adsorbed on the column and the $^{47}$Ca parent remained in solution, resulting in a 10 mL solution of 320 kBq (8.6 uCi) of $^{47}$Ca. This solution was used to produce samples for LSC and HPGe measurements. About 2 mL of additional 4 M HCl was used rinse the column and remove any residual $^{47}$Ca from the resin. Then, two 1 mL fractions of 0.1 M HCl solution were used to remove the purified $^{47}$Sc adsorbed on the column. The second of these 0.1 M HCl fractions, containing 250 kBq (6.7 uCi) of $^{47}$Sc, was used to produce a sample for LSC.

*LSC measurements:*

Liquid scintillation counting was used to verify the half-lives of both $^{47}$Ca and $^{47}$Sc as these values are crucial to the method used in this work to measure the gamma-ray branching ratios for $^{47}$Ca. These measurements were performed with a Perkin Elmer Tri-Carb 4910 TR for three $^{47}$Ca samples of different concentrations and a single $^{47}$Sc sample. Each sample was prepared with 10 mL of scintillation cocktail (Optiphase HiSafe 3, Perkin Elmer) and a small spike of activity. The three $^{47}$Ca samples (samples 1-3) were prepared with 100, 50, and 10 µL spike of the 4 M HCl solution after it was passed over the DGA resin, producing samples with 3.2, 1.6, and 0.32 kBq of $^{47}$Ca, respectively. This range of activities was used to help ensure that spectra were collected with sufficient counts without saturating the detector. Due to the smaller volume and higher activity of the purified $^{47}$Sc, a 20-times dilution was first made with the $^{47}$Sc solution, and then 50 µL of this diluted solution, containing about 0.5 kBq of $^{47}$Sc, was added to 10 mL of scintillation cocktail.

These samples were measured for 30 minutes each multiple times a day over the course of 21.7 days. "Blank" vials containing 10 mL of scintillation cocktail were measured before and after the samples at each time point in order to obtain background measurements. Additional measurements were made at 62, 93, and 140 days after the first measurement to observe any long-lived radionuclides in the samples. The LSC spectra were summed over two different energy windows: a full spectrum window and a narrowed window covering the range between the highest endpoint beta energy of $^{47}$Sc and the highest endpoint beta energy of $^{47}$Ca decay. The full spectrum window was used to count the purified $^{47}$Sc sample, as no major contaminants were observed in the spectrum. The narrowed window was selected for the $^{47}$Ca samples in order to exclude counts from $^{47}$Sc. It was verified that only background count rates were detected for the purified $^{47}$Sc sample in the narrowed window, indicating that it was an appropriate window for counting $^{47}$Ca. The uncertainty in each measurement was taken as purely statistical. Spectrum sums were background-corrected by subtracting the average measurements in the blanks at the same time point.

Origin Pro 9 software was used to fit the set of values considered for both $^{47}$Ca and $^{47}$Sc with user defined functions. For the $^{47}$Ca samples, the background-corrected count rates were fit with a simple exponential function as shown in Equation 1:

$$C(t) = C_0 e^{-\lambda t} \tag{1}$$

where $C(t)$ is the total background-corrected count rate measured in the narrowed window at time $t$ from the first spectrum collected, $C_0$ is a fitted variable for the total background-corrected count rate at time $t = 0$, and $\lambda$ is the fitted decay constant for $^{47}$Ca. No background variables were included as the longest time point spectra produced count rates equivalent to the background, indicating there was no detectable long-lived contaminant in the $^{47}$Ca samples at this range.

For the $^{47}$Sc sample, the background-corrected rates for the full spectrum window was fit with a two-part exponential function given in Equation 2:

$$D(t) = D_0 e^{-\lambda_1 t} + B_0 e^{-\lambda_2 t} \tag{2}$$

where $D(t)$ is the total background-corrected count rate measured at time $t$ from the first spectrum collected; $D_0$ and $B_0$ are fitted variables representing the initial count rates due to $^{47}$Sc and a longer-lived contaminant, respectively; and $\lambda_1$ and $\lambda_2$ are the fitted decay constants for $^{47}$Sc and the longer-lived contaminant. The contamination was reasonably attributed to a miniscule amount of $^{46}$Sc which remained after the initial purification of $^{47}$Ca, but due to the low level it was not identified in HPGe spectra.

*HPGe Gamma-Ray Spectroscopy Measurement:*

The remaining 4 M HCl solution of $^{47}$Ca that was passed through the DGA resin (300 kBq or 8.5 uCi of $^{47}$Ca) was measured in a 25 mL plastic scintillation vial with an HPGe detector (Canberra BEGe Gamma-ray Detector, BE2020) multiple times each day over the course of 21.9 days. Spectra were recorded for 30 minutes at each time point with the sample 25 cm from the face of the detector. The sample was not moved over the entire period of time that spectra were collected to preserve the exact spatial relationship between the sample and the detector face. A few long spectra were taken at 43 and 45 days from the first spectrum collected to ensure there were no long-lived contaminant radionuclides in the sample and the peaks followed the decay of $^{47}$Ca at long time points. The HPGe detector was previously energy calibrated with a $^{152}$Eu point source. Following the measurement, a 38.78 kBq $^{152}$Eu source suspended in epoxy (Eckert and Ziegler, reference date: April 1 2020, Calibrated, NIST Traceable source; 1 g/cm$^3$ epoxy; 20 mL fill volume) in an identical scintillation vial as the sample was measured in was used to calibrate the efficiency of the detector at 25 cm from the detector face.

The determination of the gamma-ray branching ratios relied on the ingrowth of $^{47}$Sc in this $^{47}$Ca sample, and accurate values for the half-life for $^{47}$Sc, a characteristic gamma-ray branching ratio for $^{47}$Sc (*i.e.*, the branching ratio of 68.3(4)% for the 159.4 keV gamma ray), and the half-life of $^{47}$Ca. With this parent-daughter relationship and a gamma spectrum of the sample, the Bateman equation can be used to quantify $^{47}$Ca in the solution at t=0, which corresponds to the time of the separation of $^{47}$Ca and $^{47}$Sc on DGA resin. The activity of $^{47}$Ca at t=0 can then be decay corrected to the time that the gamma spectrum was taken. Altogether, the use of the Bateman equation and the decay correction can be summarized in Equation 3:

$$A_1(t) = A_1(0)e^{-\lambda_1 t} = \left(\frac{\lambda_2 - \lambda_1}{\lambda_2}\right)\frac{A_2(t) - A_2(0)e^{-\lambda_2 t}}{e^{-\lambda_1 t} - e^{-\lambda_2 t}} e^{-\lambda_1 t} \qquad (3)$$

where $A_1(t)$ and $A_2(t)$ are the activity of $^{47}$Ca and $^{47}$Sc at time t, respectively; $\lambda_1$ and $\lambda_2$ are the decay constants of $^{47}$Ca and $^{47}$Sc, respectively; and time is measured from the point at which $^{47}$Ca and $^{47}$Sc were separated (meaning $A_2(0) = 0$).

The gamma spectra at each time point also provides the count rate for the three highest intensity characteristic gamma rays for $^{47}$Ca. These count rates can be adjusted by the efficiency of the detector at each energy to find the emission rate from the sample. The branching ratio for each characteristic gamma ray can then be found by taking the ratio of the emission rate for each gamma ray and the activity of $^{47}$Ca for each spectrum.

Additionally, the counts integrated in each of the three most intense $^{47}$Ca peaks and the 159.4 keV peak for $^{47}$Sc were plotted over time to verify the half-life with which these peaks decay. For the $^{47}$Ca peaks, a simple exponential, like that shown in Equation 1, was used to find the half-life for each of the three gamma rays. Since the decay of $^{47}$Ca and $^{47}$Sc affect the activity of $^{47}$Sc, the Bateman equation was used to fit these data points with the initial activity of $^{47}$Sc fixed at zero. The half-lives of $^{47}$Ca and $^{47}$Sc in addition to the initial activity of $^{47}$Ca were allowed to fit in order to find an independent measurement of the half-life of $^{47}$Sc. Ensuring that each peak sum decayed according to the correct half-life verified that no gamma rays from other background or contaminant radionuclides contributed significantly to the characteristic gamma rays of interest.

**Error Budget**

For the total count rate considered for each LSC measurement and the integrated peak sum for each HPGe measurement, the uncertainty in the measurement resulted from counting statistics (*i.e.*, the square root of the number of counts.) For the activity of $^{47}$Sc that was found from the 159.4 keV peak and used to find the half-life of $^{47}$Ca and $^{47}$Sc with the Bateman equation, only the uncertainty from the counting statistics was used as the uncertainty in each activity point in the fitting algorithm. Additional uncertainties that were common across all of the $^{47}$Sc activity data points were considered after the half-life was found in this way. The uncertainty in the 159.4 keV branching ratio and efficiency were used one at a time to change the $^{47}$Sc activity values used in the fitting routine. For example, the branching ratio for the 159.4 keV gamma ray was increased by one sigma from 68.3% to 68.7%. This new branching ratio value was used to find the activity of $^{47}$Sc, and the fitting algorithm was run again with these new activity values. This process was then repeated by setting the branching ratio back to 68.3% and changing the efficiency by one sigma. Similarly, the times associated with each data point were varied up or down by one sigma and were used in the fit algorithm. The difference between the half-lives found with each of these additional fits as the constants were varied by one sigma were added in quadrature with the error in the first half-life values for $^{47}$Ca and $^{47}$Sc found with the true branching ratio, efficiency, and times.

The uncertainty in each of the three $^{47}$Ca gamma-ray branching ratios was found in a similar way as described above for the $^{47}$Ca and $^{47}$Sc half-lives with the Bateman equation. The branching ratios were calculated with only the counting statistics for each gamma-ray peak as the uncertainty. For example, the counting statistics for the 159.4 keV peak and the 489.2 keV peak were the only uncertainties propagated through the calculation to find the 489.2 keV branching ratios. The weighted average and associated uncertainty for each gamma-ray branching ratio was found using the errors resulting from these counting statistic uncertainties. Additional uncertainties that were common across all data points were used to change their associated constants by one sigma, as explained previously, to observe the effect on the branching ratio weighted average. These uncertainties were related to the efficiency of the detector at each energy, the decay constants of $^{47}$Ca and $^{47}$Sc, the 159.4 keV branching ratio, and exact time of separation of $^{47}$Ca and $^{47}$Sc. The difference between these new weighted averages and the original weighted average was assigned as the uncertainty introduced by each constant used to find the branching ratios. A total uncertainty was calculated by adding the square of all these uncertainties and the uncertainty in the original weighted average in quadrature. Each of the uncertainties and their sources are explicitly given in the Results and Discussion section.

## Results and Discussion

Since the determination of the three $^{47}$Ca branching ratios requires the use of the half-life of $^{47}$Ca and $^{47}$Sc, these half-lives were measured with LSC. The sample used for $^{47}$Sc was a purified fraction that contained levels of $^{47}$Ca below the limit of detection using HPGe measurements, so the full spectrum window for the LSC spectra was considered for the half-life measurement of $^{47}$Sc. As $^{47}$Ca decays continuously to $^{47}$Sc, producing a purified $^{47}$Ca fraction containing no other radionuclides over a long period of time was not possible. Therefore, the channel window considered for the $^{47}$Ca half-life measurement was restricted to a region containing only signals from the decay of $^{47}$Ca.

A portion of the earliest LSC data points collected for three $^{47}$Ca samples was not considered when determining the half-life of $^{47}$Ca due to random coincidences events in these highest activity data points. The number of events in an LSC spectrum that result from random coincidences between emission from $^{47}$Ca alone has been shown to be proportional to the square of the activity of $^{47}$Ca [12]. In the same study, it was also shown that the number of random coincidences between $^{47}$Sc and $^{47}$Ca emissions is proportional to the product of the activities of the two radionuclides. Since these effects are most prominent at higher activities, the earliest time points for each of the samples were systematically removed and the resulting half-life for each sample approached the accepted $^{47}$Ca half-life value. The maximum number of data points that produced a half-life within one standard deviation of the accepted half-life value were used in the verification (*i.e.*, data points 17-41, 11-40 and 3-40 were used for samples 1-3, respectively). These data points and the fitted decay curves for each sample are shown in Figure 2a and the resulting half-lives are given in Table 1.

The $^{47}$Sc sample measured with LSC was not influenced by random coincidences to the same extent as the $^{47}$Ca samples due to a lower activity in the sample. Therefore, all the LSC measurements collected for the $^{47}$Sc sample were used to verify the half-life of $^{47}$Sc as shown in Figure 2b. The half-life found by fitting these points with Equation 2 is given in Table 1 and is only 0.03% different than the accepted half-life for $^{47}$Sc. The data suggest the presence of a very small amount of a long-lived contaminant that is consistent with $^{46}$Sc at a level of 0.5 Bq ($B_0$ in equation *2*). This is essentially negligible compared the initial activity of $^{47}$Sc in the sample of 500 Bq but became apparent at the very late time-point measurements. The half-lives of $^{47}$Ca and $^{47}$Sc measured with LSC support the accuracy of the accepted

$^{47}$Ca and $^{47}$Sc half-lives. Therefore, the accepted values of 4.536(6) and 3.3492(6) days, respectively, were used to find the three $^{47}$Ca branching ratios [1].

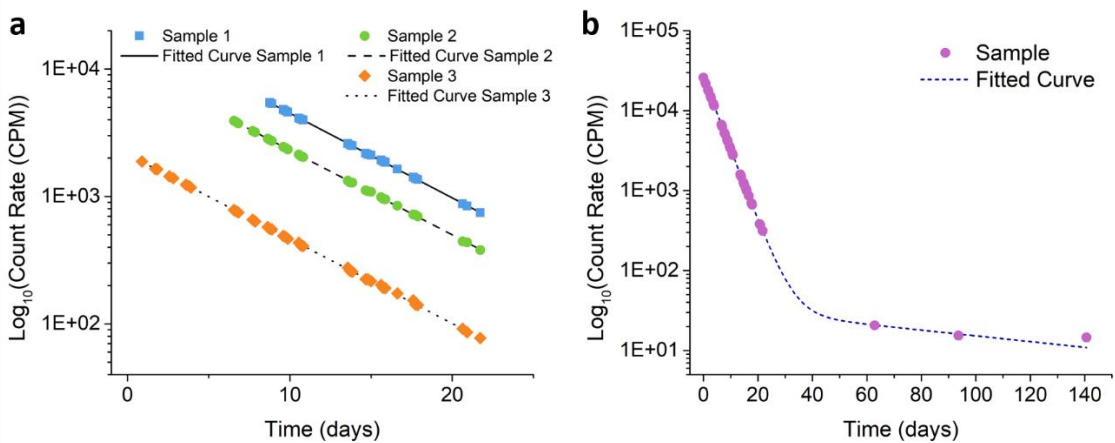

**Figure 2: Half-lives for $^{47}$Ca: LSC Count Rates**
The LSC count rates are plotted for the data points used to find the half-life of $^{47}$Ca (a) and $^{47}$Sc (b). The fitted decay curves found for each sample are overlaid. Error bars found from counting statistics are given for each point but are largely not visible as they are smaller than the size of the plotted points.

The half-lives with which the 489.2, 807.9, 1297.1, and 159.4 keV gamma-ray peaks decayed across the gamma spectra for the $^{47}$Ca/$^{47}$Sc sample were also found to verify the purity of these gamma-ray peaks. Figure 3a gives the integrated peak counts for the three $^{47}$Ca gamma rays and a simple exponential fitted function for each peak. Since the baseline corrected integrated counts were used in this fit, only uncertainties from counting statistics and baseline corrections for each data point were considered. The activity of $^{47}$Sc resulting from the integrated counts in the 159.4 keV peak across the gamma spectra recorded is given in Figure 3b. The Bateman equation was used to fit the 159 keV data points first with the half-life of both radionuclides and the initial activity of $^{47}$Ca as variables and again with only the half-life of $^{47}$Sc and the initial activity of $^{47}$Ca as variables. Only uncertainties from the baseline correction and counting statistics were used in both fitting routines. As described previously, the uncertainties in the constants used to find the activity of $^{47}$Sc for each spectrum were systematically varied to find the uncertainty contribution from each constant (Table 2). With the dead time of the detector at 1% or less for each spectrum in this analysis, a dead time correction was not necessary for accurate results.

The four resulting values for the $^{47}$Ca half-life and two for the $^{47}$Sc half-life found in this way are given in Table 1. Each measurement agrees quite well with the corresponding evaluated value with the error bar for the half-lives found using the Bateman equation being larger, 3% for the half-life of $^{47}$Sc and 2% for the half-life of $^{47}$Ca, when these values were both used as variables. When the half-life of $^{47}$Ca was fixed at the accepted value, however, the half-life of $^{47}$Sc was found with only a 0.2% uncertainty. These results confirm that, as expected, the 489.2, 807.9, and 1297.1 keV gamma rays come from $^{47}$Ca and the 159.4 keV gamma ray comes from $^{47}$Sc with minor to no interferences from other sources.

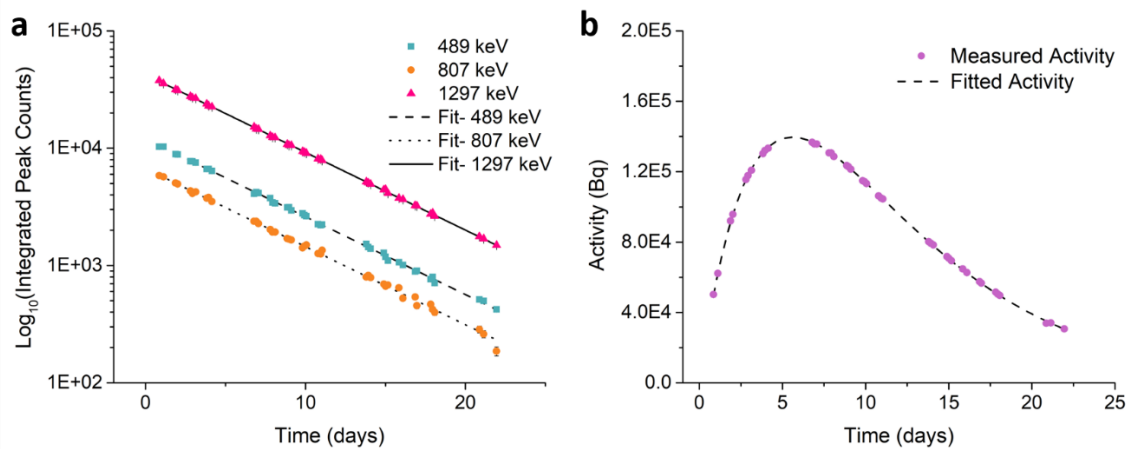

**Figure 3: Half-lives for $^{47}$Ca and $^{47}$Sc: Gamma Spectroscopy Integrated Peaks**
The integrated peak counts for the 489.2, 807.9, and 1297.1 keV gamma ray for $^{47}$Ca (a) and the $^{47}$Sc activity found with the 159.4 keV gamma ray (b) are given for each gamma spectrum. These data points are fit with a simple exponential for the $^{47}$Ca gamma rays and with the Bateman equation for the $^{47}$Sc activity. Error bars are given from counting statistics and baseline corrections and are largely not visible outside of the plotted data points.

**Table 1. Half-lives of $^{47}$Ca and $^{47}$Sc**
The reduced chi-squared value is given to provide information on the goodness of fit for value found. These values are the chi-squared value divided by the degrees of freedom given in parenthesis after the value.

| Radionuclide | Source | Half-life (days) | Reduced Chi-squared (DOF) |
|---|---|---|---|
| $^{47}$Ca | **Evaluated half-life [1]** | **4.536(3)** | - |
| | LSC, sample 1 | 4.53(1) | 2.57 (23) |
| | LSC, sample 2 | 4.532(9) | 1.90 (28) |
| | LSC, sample 3 | 4.53(1) | 1.56 (36) |
| | Average of LSC measurements | 4.531(5) | - |
| | HPGe, 489 keV gamma ray | 4.54(2) | 1.38 (39) |
| | HPGe, 807 keV gamma ray | 4.52(3) | 1.35 (39) |
| | HPGe, 1297 keV gamma ray | 4.53(1) | 1.19 (39) |
| | Average HPGe direct measurement | 4.534(3) | - |
| | HPGe, 159 keV gamma ray | 4.6(1) | 1.07 (38) |
| $^{47}$Sc | **Evaluated half-life [1]** | **3.3492(6)** | - |
| | LSC | 3.350(2) | 1.03 (39) |
| | HPGe- 159 keV gamma ray | 3.3(1) | 1.07 (38) |
| | HPGe- 159 keV gamma ray; $^{47}$Ca decay constant fixed | 3.349(8) | 3.91 (39) |

**Table 2. Error Budget for Half-lives Found from 159.4 keV Gamma-Ray Peak**

The uncertainty related to the separation time resulted from the approximately 8 minutes required to separate $^{47}$Ca and $^{47}$Sc on the DGA resin. The separation time was chosen as the midpoint of this separation length and the uncertainty was assigned as half of the separation length. A dash in the table indicates that the source of uncertainty was not relevant for the measurement whereas a value of zero indicates that the source of uncertainty was considered for the measurement but resulted in no additional uncertainty. In particular, the efficiency and branching ratio for 159.4 keV were constants with uncertainties relevant for finding the half-life of $^{47}$Sc. However, these uncertainties resulted in no change in the fitted half-life since they uniformly shift all data points and do not affect the decay rate.

| Source of Uncertainty | Half-life | | |
|---|---|---|---|
| | $^{47}$Ca | $^{47}$Sc | $^{47}$Sc with Constant $^{47}$Ca Half-life |
| Counting Statistics and Baseline Correction | 0.084 | 0.066 | 0.0067 |
| Efficiency at 159.4 keV | 0.040 | 0.024 | 0.0000 |
| Branching Ratio- 159.4 keV | 0.063 | 0.044 | 0.0000 |
| Separation Time | 0.085 | 0.064 | 0.0032 |
| Decay constant of $^{47}$Ca | - | - | 0.0026 |
| **Total** | **0.14** | **0.10** | **0.0078** |

With confirmation that both sets of half-lives for $^{47}$Ca and $^{47}$Sc agree with the evaluated half-lives, the gamma spectra of the $^{47}$Ca/$^{47}$Sc sample were analyzed for the branching ratios of the three main $^{47}$Ca gamma rays. The branching ratios for each gamma spectrum are given for the 489.2 and 807.9 keV gamma ray in Figure 4a and for the 1297.1 keV gamma ray in Figure 4b. The average value across each set of data points is given as a solid or dashed line through the points. The error bars in Figure 4a and b result from the counting statistics from each integrated peak used to find the branching ratio values as these are the only uncertainties that vary by data point. The other errors that influence the branching ratio value were incorporated in the final error of the weighted average value. All errors considered for these measurements are given in Table 3 where they are listed by the source of the uncertainty. It is also vital to note for this measurement that the emission of the 489.2 and 807.9 keV gamma rays in the cascade from the 1297.1 keV excited state is isotropic with angular correlation coefficients of $A_2 = -0.050(6)$ and $A_4 = 0$ [9,13]. With this isotropic distribution, the probability of simultaneously detecting both the 489 and 807 keV emission from the same decay event is on the order of the detection efficiency for these gamma rays (*i.e.*, 1E-4 for 807.9 keV and 3E-4 for 489.2 keV). Since this probability is small compared to the uncertainties in these measurements, no correction was made to the branching ratios for this coincidence probability.

The final values found for the branching ratios of the 489.2, 807.9 and 1297.1 keV gamma rays are 7.17(5)%, 7.11(5)%, and 75.0(5)%, respectively. These values differ significantly from the current evaluated branching ratios of these gamma rays, varying about 20% for 489.2 and 807.9 keV and 12% for 1297.1 keV. Additionally, the precision of these branching ratios is greatly increased over the evaluated values with uncertainties of less than 1% for each branching ratio.

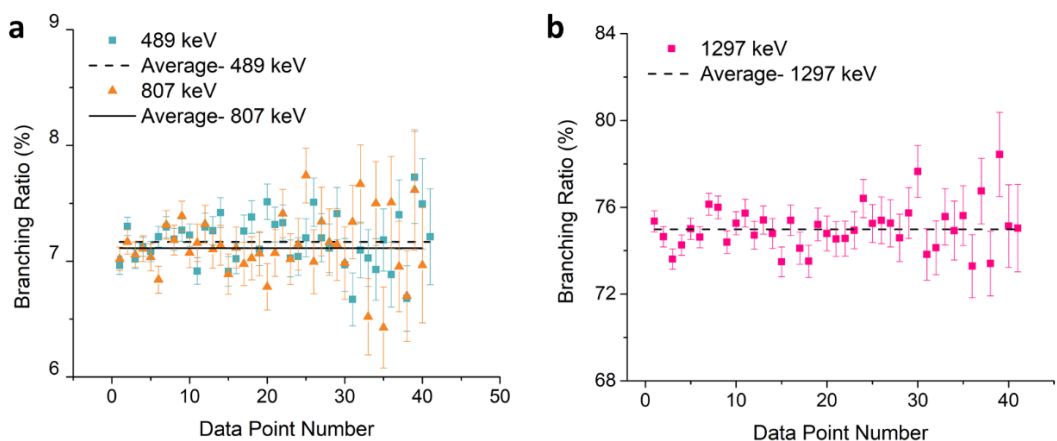

**Figure 4: Branching Ratios for Three Most Intense $^{47}$Ca Gamma Rays**
The branching ratios for each gamma spectrum considered are given for the 489.2 and 807.9 keV gamma ray (a) and the 1297.1 keV gamma ray (b). The solid or dashed line through the data points indicates the average branching ratio for each gamma-ray energy.

**Table 3. Error Budget for Three $^{47}$Ca Branching Ratios**
The absolute uncertainty for the branching ratio per 100 decays is given for different sources of uncertainty. The uncertainty related to the separation time resulted from the approximately 8 minutes required to separate $^{47}$Ca and $^{47}$Sc on the DGA resin. The separation time was chosen as the midpoint of this separation length and the uncertainty was assigned as half of the separation length. Additionally, the efficiency uncertainty was found by varying the uncertainty at 159.4 keV and one $^{47}$Ca gamma-ray energy at the same time since these uncertainties were related.

| Source of Uncertainty | Energy of Gamma Ray (keV) | | |
| --- | --- | --- | --- |
|  | 489.2 | 807.9 | 1297.1 |
| Counting Statistics | 2.06E-2 | 2.78E-2 | 1.13E-1 |
| Efficiency | 1.84E-2 | 1.83E-2 | 5.85E-2 |
| Decay Constant- $^{47}$Ca | 1.97E-3 | 1.92E-3 | 2.18E-2 |
| Decay Constant- $^{47}$Sc | 5.6E-4 | 5.7E-4 | 5.4E-3 |
| Branching Ratio- 159.4 keV | 4.2E-2 | 4.2E-2 | 4.4E-1 |
| Separation Time | 7.0E-3 | 7.1E-3 | 2.6.7E-2 |
| **Total** | **5.1E-2** | **5.4E-2** | **4.6E-1** |

With these remeasured branching ratios for $^{47}$Ca, other minor gamma-ray branching ratios can be updated in the $^{47}$Ca decay scheme (see Figure 5). In particular, the 41.1, 530.6, 767.1, and 1878 keV gamma rays have been measured previously through a ratio with the more intense gamma rays in the $^{47}$Ca decay scheme. Additionally, the 731.6 and 1147 keV gamma rays have been inferred through intensity balancing from the 1878 keV excited state. The branching ratio as an emission percent for $^{47}$Ca decays based on the 489.2, 807.9, and 1297.1 keV branching ratios presented in this work are given in Table 4 and Figure 5.

**Table 4. Minor Gamma Ray Branching Ratio Values**

| Gamma-Ray Energy (keV) | Measurement Description | Ratio | Evaluated Branching Ratio (%) | Branching Ratio from This Work (%) |
|---|---|---|---|---|
| 41.1 | Relative to intensity 100 for 1297.1 keV gamma ray [14] | 0.0085(10) | 0.0056(13) | 0.0064(8) |
| 530.6 | Ratio with 489.2 keV gamma ray [13] | 0.0146(10) | 0.086(18) | 0.105(7) |
| 731.6 | Intensity balanced 1878 keV energy level [1] | - | 0.011(3) | 0.009(2) |
| 767.1 | Ratio with 807.9 keV gamma ray [13] | 0.0294(10) | 0.18(4) | 0.209(7) |
| 1147 | Intensity balanced 1878 keV energy level [1] | - | 0.011(3) | 0.009(2) |
| 1878 | Relative to intensity 100 for 1297.1 keV gamma ray [14] | 0.038(4) | 0.025(6) | 0.028(3) |

The beta decay from the ground state of $^{47}$Ca to both the ground state of $^{47}$Sc and the 1297.1 keV excited state of $^{47}$Sc are currently reported with high uncertainties. These values can both be found with far lower uncertainty with the 489.2 and 1297.1 keV branching ratios reported herein. The relationship between key nuclear data in the decay of $^{47}$Ca is shown in the decay scheme in Figure 5. Since only the 489.2, 530.6 and 1297.1 keV gamma rays are emitted from the 1297.1 keV excited state for $^{47}$Sc, the beta decay branching ratio is equal to the sum of the branching ratios of these three gamma rays. Internal conversion does occur to de-excite from this energy level, but the rates are negligible compared to even the small uncertainties in these three gamma-ray branching ratios [1]. With the branching ratios measured in this work and updated 530.6 keV branching ratio given in Table 4, the beta decay branching ratio from the ground state of $^{47}$Ca to the 1297.1 keV excited state of $^{47}$Sc is found to be 82.2(5)% [13]. This updated beta decay branching ratio can be used with the previously reported branching ratios for beta decay from the ground state of $^{47}$Ca to the 766.8 and 1878.2 keV excited states of $^{47}$Sc (*i.e.*, 0.087(3) and 0.037(8)%, respectively) [1,13–15]. Intensity balancing with these values gives a beta decay branching ratio of 17.7(5)% for the ground state to ground state beta decay of $^{47}$Ca to $^{47}$Sc. The branching ratios for the beta decay of $^{47}$Ca to the 1297.1 keV excited state and the ground state of $^{47}$Sc given here fall within the large uncertainty of the accepted values but are much more precise. The values for all four beta decay branches are given in the decay scheme in Figure 5.

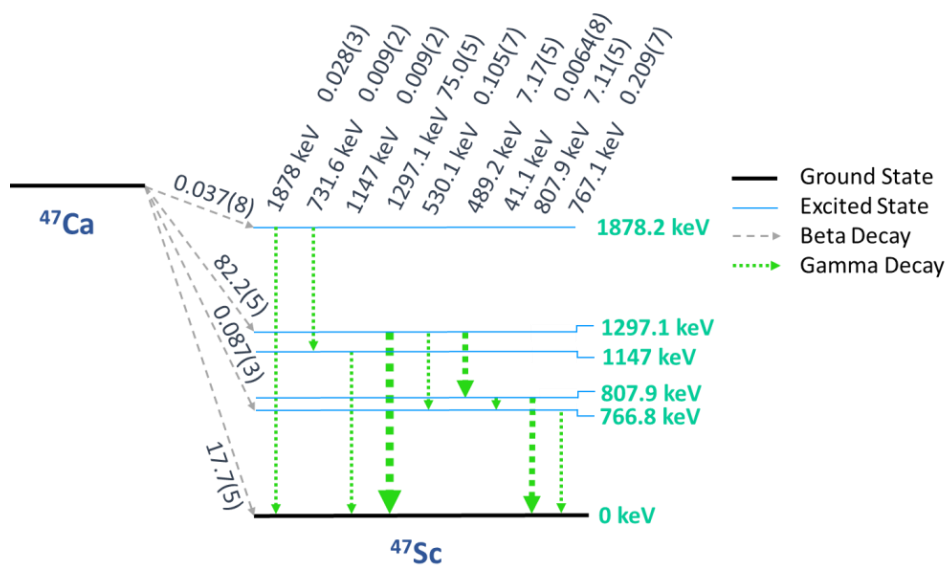

**Figure 5: Decay Scheme of ⁴⁷Ca Beta Decay to ⁴⁷Sc**
The four beta decay branches for ⁴⁷Ca and the subsequent gamma-ray emissions from the excited energy levels in ⁴⁷Sc are given. The 489.2, 530.6, and 1297.1 keV gamma rays are the dominant emissions from the 1297.1 keV excited state in ⁴⁷Sc. The branching ratios for each beta and gamma-ray emission branches are indicated in the figure as a percent.

## Conclusion

With a sample of ⁴⁷Ca produced at the NSCL, the half-life of ⁴⁷Ca and its daughter, ⁴⁷Sc, were measured with LSC and HPGe gamma-ray spectra and were found to agree with the current evaluated half-lives. Using the ingrowth of ⁴⁷Sc and well-known nuclear data for this radionuclide, the branching ratio for the 489.2, 807.9, and 1297.1 keV ⁴⁷Ca gamma rays were remeasured as 7.17(5)%, 7.11(5)%, and 75.0(5)%, respectively. These values are more precise than the current accepted branching ratios for these gamma rays. Additionally, the beta decay branching ratio from the ground state of ⁴⁷Ca to the 1297.1 excited state and the ground state of ⁴⁷Sc were found to be 82.2(5)% and 17.7(5)%, respectively. Although these values fall within the uncertainty range for the currently reported values, they have a marked increase in precision. These updated values will allow for more accurate and precise quantification of ⁴⁷Ca when using ⁴⁷Ca to generate ⁴⁷Sc for nuclear medicine applications.

## Acknowledgements

The authors would like to thank the beam operators and the A1900 group for providing the accelerated beam used to produce ⁴⁷Ca and Dirk Weisshaar for his assistance with data collection and processing. The time that Katharina Domnanich, Chirag Vyas, Wes Walker, Scott Essenmacher, Samridhi Satija, and Hannah Clause dedicated to the ⁴⁸Ca beam experiment that produced this ⁴⁷Ca sample is also greatly appreciated. This material is based upon work supported by the U.S. Department of Energy, Office of Science, Office of Nuclear Physics, DOE Isotope Program, under a funded proposal from DOE-FOA-0001588 (DE-SC0018637). Additional support was provided by the U.S. Department of Energy NNSA SSGF Program (DE-NA0003864) and Michigan State University.

## References

[1] T. W. Burrows, *Nuclear Data Sheets for A = 47*, Nucl. Data Sheets **108**, 923 (2007).
[2] D. Reher, H. H. Hansen, R. Vaninbroukx, M. J. Woods, C. E. Grant, S. E. M. Lucas, J. Bouchard, J. Morel, and R. Vatin, *The Decay of 47Sc*, Appl. Radiat. Isot. **37**, 973 (1986).
[3] J. W. T. Meadows and V. A. Mode, *A REDETERMINATION OF THE OF SCANDIUM-47*, J. Inorg. Nucl. Chem. **30**, 361 (1968).
[4] H. Mommsen, I. Perlman, and J. Yellin, *47Sc HalfLife*, Nucl. Instruments Methods **177**, 545 (1980).
[5] K. A. Domnanich, C. Müller, M. Benešová, R. Dressler, S. Haller, U. Köster, B. Ponsard, R. Schibli, A. Türler, and N. P. van der Meulen, *47Sc as Useful β−-Emitter for the Radiotheragnostic Paradigm: A Comparative Study of Feasible Production Routes*, EJNMMI Radiopharm. Chem. **2**, 5 (2017).
[6] C. Müller, M. Bunka, S. Haller, U. Köster, N. van der Meulen, A. Türler, and R. Schibli, *Preclinical Application of 47Sc-Folate – A Pilot Study in Tumor-Bearing Mice*, Nucl. Med. Biol. **41**, (2014).
[7] C. Müller, M. Bunka, S. Haller, U. Köster, V. Groehn, P. Bernhardt, N. van der Meulen, A. Türler, and R. Schibli, *Promising Prospects for 44Sc-/47Sc-Based Theragnostics: Application of 47Sc for Radionuclide Tumor Therapy in Mice.*, J. Nucl. Med. **55**, 1658 (2014).
[8] C. Müller, K. A. Domnanich, C. A. Umbricht, and N. P. Van Der Meulen, *Scandium and Terbium Radionuclides for Radiotheranostics: Current State of Development towards Clinical Application*, Br. J. Radiol. **91**, (2018).
[9] M. B. Lewis, *Nuclear Data Sheets for A = 47*, Nucl. Data Sheets 313 (1970).
[10] K. A. Domnanich, E. P. Abel, H. K. Clause, C. Kalman, W. Walker, and G. W. Severin, *An Isotope Harvesting Beam Blocker for the National Superconducting Cyclotron Laboratory*, Nucl. Instruments Methods Phys. Res. Sect. A Accel. Spectrometers, Detect. Assoc. Equip. **959**, 163526 (2020).
[11] E. P. Abel, K. Domnanich, H. K. Clause, C. Kalman, W. Walker, J. A. Shusterman, J. Greene, M. Gott, and G. W. Severin, *Production, Collection, and Purification of 47 Ca for the Generation of 47 Sc through Isotope Harvesting at the National Superconducting Cyclotron Laboratory*, ACS Omega (2020).
[12] L. Burkinshaw, D. H. Marshall, and C. B. Oxby, *The Half-Life of Calcium-47*, Int. J. Appl. Radiat. Isot. **20**, 393 (1969).
[13] M. S. Freedman, F. T. Porter, and F. Wagner, *Evidence for Hindered First-Forbidden Unique Beta Branch in Ca-47 Decay*, Phys. Rev. **152**, 1005 (1966).
[14] R. E. Wood, J. M. Palms, and P. V. Rao, *Note on the Gamma Ray Spectrum from the 47Ca Decay*, Nucl. Physics, Sect. A **126**, 300 (1969).
[15] H. J. Fischbeck, *Intensities of Inner Beta Groups in Ca-47 and Shapes of Nonstatistical Allowed Outer Groups in Ca-47 and P-32*, Phys. Rev. **173**, 1078 (1968).